\documentclass[aps,apl,twocolumn,groupedaddress]{revtex4}

\usepackage{amsmath}

\usepackage{graphics}

\usepackage{epsf}

\begin{document}
\setlength{\unitlength}{12pt}

\title{Transition from Ferromagnetism to Antiferromagnetism in Ga$_{1-x}$Mn$_x$N}

\author{Gustavo M. Dalpian and Su-Huai Wei}

\affiliation{National Renewable Energy Laboratory, Golden, Colorado 80401, USA }

\date{\today}

\begin{abstract}
Using density functional theory, we study the magnetic stability of the
Ga$_{1-x}$Mn$_x$N alloy system. We show that unlike Ga$_{1-x}$Mn$_x$As, which
shows only ferromagnetic (FM) phase, Ga$_{1-x}$Mn$_x$N can be stable in either
FM or antiferromagnetic phases depending on the alloy concentration. The magnetic order can also be altered by applying
pressure or with charge compensation. A unified model is used to explain these
behaviors.
\end{abstract}


\maketitle


The discovery of ferromagnetism in Mn-containing III-V semiconductors
has attracted significant attention in the last decade because the
interesting combination of semiconductor electronics and metallic
ferromagnetism creates great potential for developing functional
devices that manipulate spin, as well as charge. Among the materials
that have been investigated, GaMnN is one of the most interesting
systems. On one hand, original theoretical predictions suggest that high
$T_c$ ferromagnetism can be achieved in the GaMnN alloy \cite{dietl},
which has resulted in an extensive experimental, as well as theoretical,
study of its physical properties. On the other hand, available
experimental data often contradict each other. Some reports
show that high $T_c$ ferromagnetism is achievable in this
system \cite{over01,reed01,theo01,shon02}; others show that the
magnetic coupling of the Mn ions in GaMnN is actually antiferromagnetic
(AFM) \cite{zajac01}. The exact nature of the magnetism observed in
this system is also still under debate \cite{puru,dhar03,pearton04}. 
Some groups suggested that the
observed ferromagnetism could be due to secondary phases generated during
growth, and that GaMnN is a spin glass system \cite{dhar03}. Other
groups argue that no secondary phase is observed in their single-crystal
GaMnN samples that show ferromagnetism \cite{pearton04}. There
are also discussions about whether the relevant Mn $3d$ state has
$d^4$ or $d^5+h$ character, or whether the $d$ levels are localized in
the bandgap or inside the valence band \cite{dietl03}.

Although many of the issues are still under intensive study, recent
{\it ab initio} band structure and total energy
calculations \cite{kron02,kula02,sanyal03} seem to agree that Mn $3d$
levels are located in the gap, and that the interaction between
substitutional Mn ions is ferromagnetic (FM) at low Mn
concentration.  Because previous {\it ab initio} studies also
find that pure MnN has an AFM ground state \cite{janotti03}, 
an interesting question was raised about how the magnetic and electronic
properties of Ga$_{1-x}$Mn$_x$N evolve as a function of the Mn
concentration $x$.  Furthermore, it is now well known that the
ferromagnetism observed in Mn-containing GaAs is caused by holes in
the host valence-band-derived states. It would be important to
understand how the mechanism changes in Mn-containing GaN, where the
holes are created in the Mn $d$ bands.

In this paper, using {\it ab initio} band structure and total energy
methods, we study the magnetic properties of Ga$_{1-x}$Mn$_x$N in the
low and high Mn concentration regimes. We also study the effects of
pressure and charge compensation on the magnetic properties of this
system. We show that unlike in GaMnAs, where $p$-$d$ coupling induced
level splitting at the valence band maximum (VBM) is the dominant effect
that stabilizes the FM phase, direct Mn-Mn $d$-$d$ coupling 
mediated by anion $p$ states is
the dominant effect in GaMnN \cite{our_prl}. We find that in GaMnN: (i) The Mn $3d$ states are
inside the band gap, in agreement with previous calculations.  (ii) At
low Mn concentrations, the Mn atom tends to have a higher magnetic moment
and the magnetic ground state is FM. (iii) At higher
concentration, the magnetic moment of Mn decreases and the ground state
changes to AFM phase. (iv) When the material is
compressed, the system becomes more stable in the AFM phase, even at
low concentrations.  (v) When Mn is negatively charged, the system
turns to AFM, even at low concentrations.  We show that all these results
can be explained by the level crossing of Mn $d$ spin-up and spin-down
states, which varies with Mn concentration and pressure.

The calculations were performed using an {\it ab initio} plane wave
basis code \cite{vasp}, based on the density functional theory and
using ultrasoft pseudopotentials \cite{uspp}.  For the exchange and
correlation potential, we used the generalized gradient approximation
of Perdew and Wang \cite{pw91}. All the structural degrees of
freedom (lattice parameters and atomic positions) are optimized by
minimizing quantum mechanical forces and total energy.  The Brillouin
zone integration is performed using the Monkhost-Pack special {\bf k}
points scheme \cite{monk76}. A large number of $k$ points and high
cut-off energies for the basis functions are used to ensure a
convergence error within a few meV.  We considered the zinc-blende
alloy and assumed that the same results also hold for the alloy in a
wurtzite structure. The disorder of the alloy is taken into account
explicitly through the special quasirandom structure
approach \cite{sqs}. Supercells up to 64 atoms have been used.
For AFM calculations involving more than two Mn
atoms, the sign of the magnetic moments on each Mn site are initially distributed
randomly. For the cases where the AFM configuration was higher in
energy than the FM phase, several other AFM magnetic configurations are
tested to make sure that the FM phase indeed has the lowest total
energy.

\begin{figure}[h]
\epsfxsize 7.5cm
\centerline{\epsffile{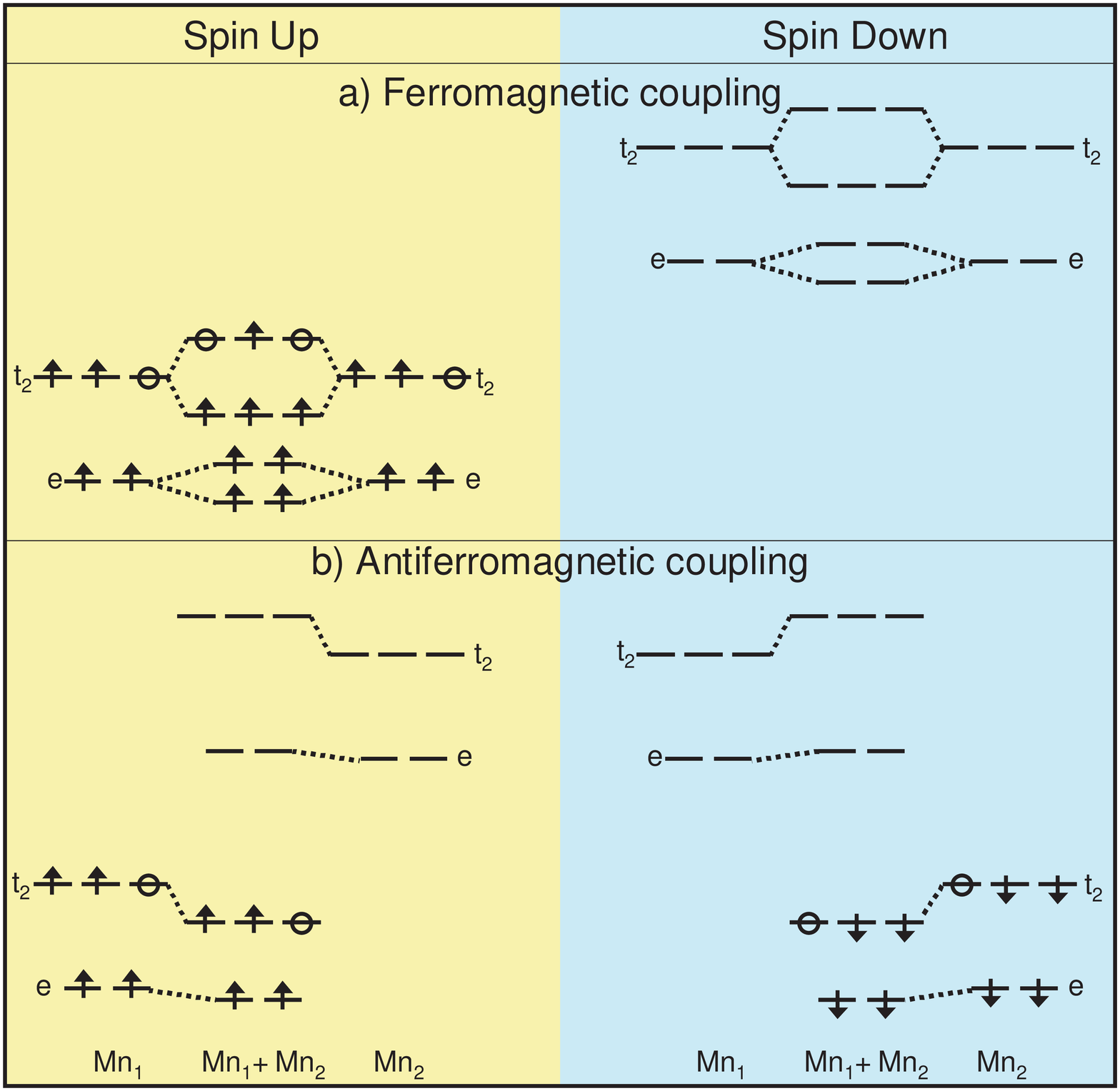}}
\caption{Schematic model for the hole-mediated stabilization of ferromagnetic phase
in Ga$_{1-x}$Mn$_x$N with low Mn concentration. In this case, the crystal field splitting is smaller
than the exchange splitting. The $t_2$ state is the hybridized state with majority Mn $t_{2d}$
characeter. For simplicity, the N $t_{2p}$ state is not shown.
\label{fig1}}
\end{figure}

We find that unlike Ga$_{1-x}$Mn$_x$As, where the ground state is always
FM, in Ga$_{1-x}$Mn$_x$N, the magnetic ground state changes with Mn
concentration $x$.  At low Mn concentrations, our calculations show that
the FM phase of the Mn atoms is more stable, in agreement with
other previous theoretical calculations \cite{sanyal03}. However, at
high Mn concentrations, the lowest energy state becomes AFM. To
understand this interesting behavior, we have drawn schematically in
Figs. 1 and 2 the Mn $d$ energy levels and the
coupling between them in the spin-up and spin down states. Because the
spin-orbit coupling is neglected in our discussion, only states
with the same spin and symmetry can couple to each other. We
find that when Mn atoms are added to GaN, the Mn $3d$ levels are
introduced inside the band gap with a $d^4$ configuration. At each spin
channel, due to the crystal field splitting, the $t_{2d}$ state is
above the $e_d$ state. The crystal field splitting increases when Mn
concentration increases due to the coupling between the Mn $t_{2d}$
and N $t_{2p}$ states \cite{note1}.

\begin{figure}[h]
\epsfxsize7.5cm
\centerline{\epsffile{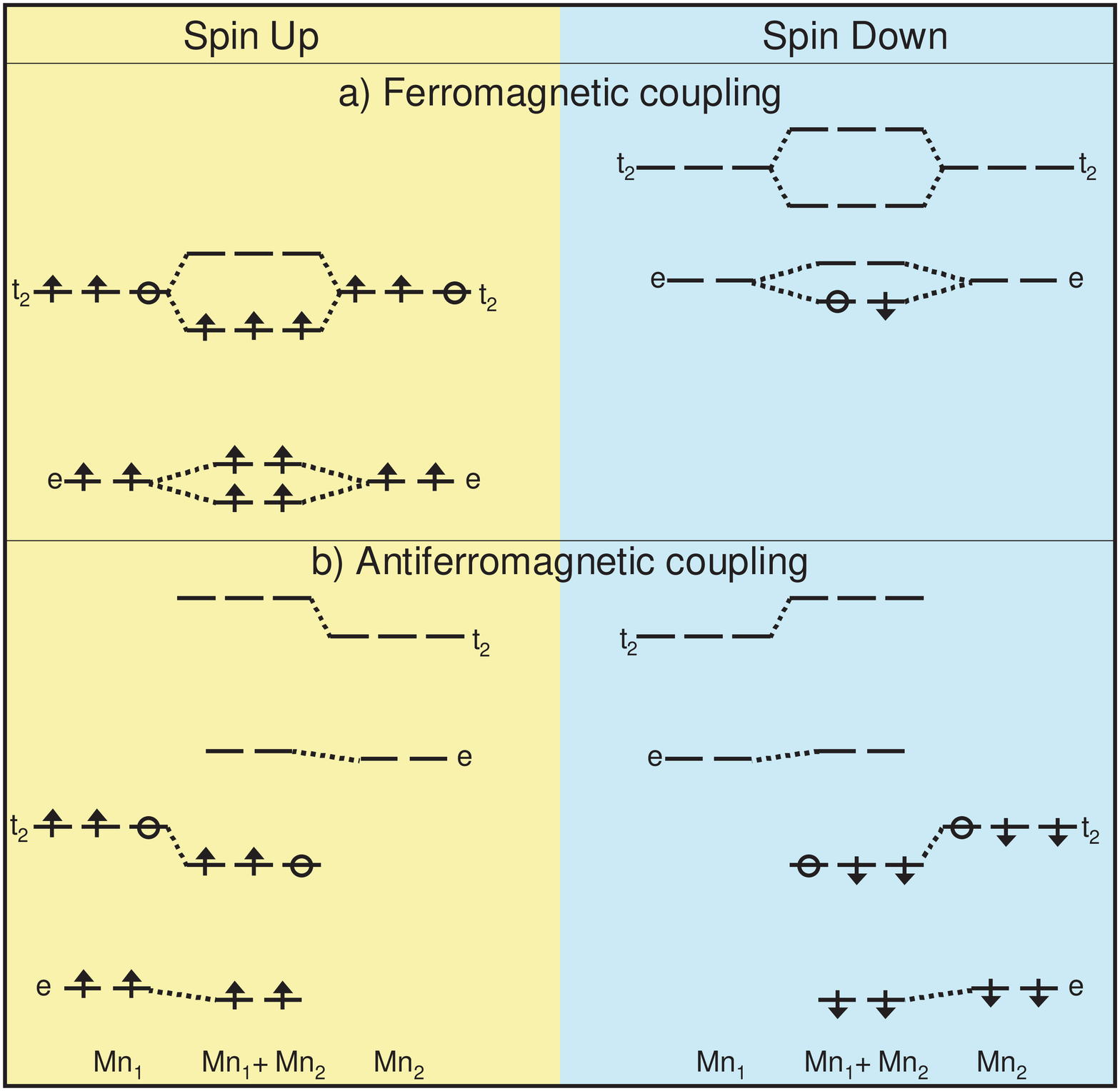}}
\caption{Schematic model for the stabilization of
antiferromagnetic phase in Ga$_{1-x}$Mn$_x$N with high Mn concentrations.  In this case, the
crystal field splitting is comparable to the exchange splitting.
\label{fig2}}
\end{figure}

When the crystal field splitting
is relatively smaller than the spin exchange splitting, the majority
spin states will be filled, except for the holes in the $t_{2d}$
state, while the minority spin states are empty (Fig. 1). In the FM
configuration, the majority spin state of neighboring Mn atoms couple
to each other (Fig. 1a), and the level repulsion pushes one level up and one
down. Because the $t_{2d}$ state is not fully occupied, the level
repulsion puts the holes in the higher energy level, and can thus
stabilize the FM phase \cite{our_prl,schilf01}.  The energy gain in this case
depends on the position of the Mn-Mn pair and increases with the hole 
concentration at the $t_{2d}$
level. In the AFM configuration the majority spin state of one Mn
atom couples only to the minority spin states of the other Mn atom
(Fig. 1b) with opposite moment. This level repulsion push the occupied
levels down and the unoccupied levels up, thus can stabilize the
AFM phase. The coupling increases when the spin exchange splitting between
the occupied and unoccupied states decreases and when the hole concentration
at the $t_{2d}$ state decreases.

\begin{figure}[h]
\epsfxsize 7.5cm
\centerline{\epsffile{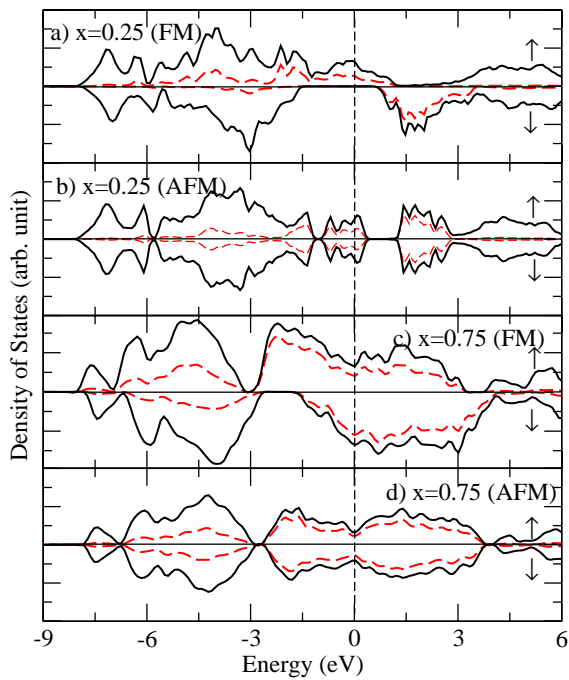}}
\caption{Total (solid) and Mn$_d$ projected (dashed) density of states
for Ga$_{1-x}$Mn$_x$N with different concentrations and magnetic
configurations. FM is for ferromagnetic and AFM for antiferromagnetic.
The Fermi energy is at zero energy.
\label{fig3}}
\end{figure}

At low Mn concentration, because the $p$-$d$ repulsion is weak, the
exchange splitting is larger than the crystal field splitting (Fig.~1) and the
FM interaction is larger than the AFM one. This explains why
GaMnN has a FM ground state at low Mn concentration.  When the Mn
concentration increases, the crystal field
splitting increases due to  the larger $p$-$d$ repulsion.
The dispersion of the Mn$_d$ band also increases. 
When part of the majority spin $t_{2d}$ levels becomes
higher than the minority spin $e_d$ state, charge transfer will
occur between these two states, which leads to a reduced spin
exchange splitting, as shown in Fig. 2. Because reduced spin exchange
splitting will enhance the AFM coupling between the Mn $d$ majority
state and the minority state, the system will become increasingly
stable in the AFM phase when the Mn concentration increases. It is
worth noting that when the crystal field splitting increases further,
only the $e_d$ states will be occupied and the system will become
paramagnetic, with no magnetic moment at the Mn site.

\begin{figure}[b]
\epsfxsize 8.5cm
\centerline{\epsffile{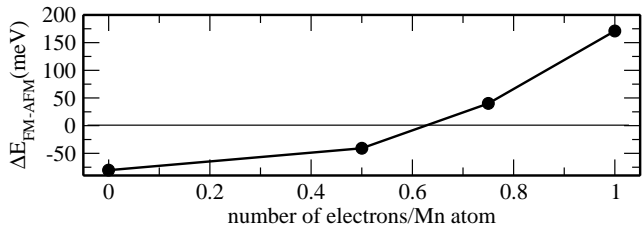}}
\caption{Energy difference between ferromagnetic and antiferromagnetic configurations as a
function of the number of electrons added per Mn atom in Ga$_{0.75}$Mn$_{0.25}$N.
\label{fig4}}
\end{figure}

To test the model discussed above, we plotted in Fig. 3 the
total and the projected density of states (DOS) of Ga$_{1-x}$Mn$_x$N
with $x=0.25$ and $x=0.75$, which has FM and AFM ground states,
respectively. From the calculated projected density of states (PDOS),
we find that at $x=0.25$, in the FM case, the holes are created in the
spin-up channel, whereas in the AFM phase, the holes are created in
both spin channels. The calculated magnetic moment is 3.59\,$\mu_B$ for
the FM phase and 3.32\,$\mu_B$ in the AFM phase. The reason that the FM
phase has a larger magnetic moment is because the AFM coupling shown in
Fig. 1b mixes filled and empty $d$ states \cite{gongprb}, thus reducing the
magnetic moment in the AFM phase.  At $x=0.75$, the increase of the Mn
concentration also increases the $p$-$d$ repulsion, leading to a large
overlap between the majority spin $t_{2d}$ levels and the minority
spin $e_d$ levels.  Due to the charge transfer between the majority
spin and minority spin states, the minority $e_d$ state is partially
occupied and the magnetic moment is reduced. In the FM and AFM phases,
the calculated magnetic moments are 1.96 and 2.40\,$\mu_B$,
respectively. In this case, the FM phase has a smaller moment than the AFM
phase, opposite to that at low Mn concentration.  This is because at
higher concentration, the charge transfer from the spin-up $t_{2d}$ level
to the spin-down $e_d$ level is larger in the FM phase than in the AFM phase (Fig.~2). We see
that the calculated results are consistent with our model.

Our discussion above shows that the change from FM to AFM in GaMnN when Mn concentration increases is due
to the increased crystal field splitting and band broadening, which leads to a reduced Mn $d$-$d$ spin
exchange splitting.  We notice that the same
effect can also be simulated by applying pressure (or reducing the lattice constant).
This is because under pressure, the increased $p$-$d$ coupling increases the
crystal field splitting.
To test this, we repeated the calculation at $x=0.25$, but at a lattice
constant that is 10\% smaller than the equilibrium lattice constant.
We find that under this compression, the system indeed becomes more stable
in the AFM phase, whereas at its equilibrium lattice constant, it is more
stable in the FM phase. The magnetic moment in this case is also reduced, being 2.48\,$\mu_B$ for the
AFM phase and 2.41\,$\mu_B$ for the FM phase, similar to the case of high concentrations.
A similar effect was previously observed in the surface of GaN, 
where the distance between
Mn atoms is smaller \cite{puru}.

The model we discussed above (Fig. 1) also shows that the FM phase is more stable when holes are
in the Mn $d$ bands, whereas the AFM phase will be more stable when the holes are filled \cite{our_prl}.
 To test this, we have 
calculated the energy difference $\Delta$E$_{FM-AFM}$ as a function of the number of
electrons added per Mn atom at $x=0.25$.
The results are plotted in Fig. 4. We find that, indeed, 
the system becomes more stable in the AFM phase when the added electron reaches 0.62 per Mn atom.
We also notice that the lattice constant increases when
Mn is negatively charged because of the larger Coulombic repulsion caused by the extra electrons.
These results confirmed our model. They also suggest that the AFM phase is possible in
GaMnN if the Mn atoms are compensated by donors such as Mn
interstitials, N vacancies, or N$_{Ga}$ antisite defects. 

In summary, we have explained the magnetic behavior
of GaMnN using a band structure model. We show that because the Mn $d$ states are inside the GaN band gap,
its behavior is quite different from that of GaMnAs. The stability of its magnetic
phases can change from FM to AFM, depending on the Mn concentration, pressure, and charge
compensation.

This work is funded by the U.S. Department of Energy, 
Office of Science, Basic Energy Sciences, under Contract No. DE-AC36-99GO10337 to NREL.

\newpage


\begin{thebibliography}{19}
\bibitem[{not()}]{note1}
\bibinfo{note}{We find that because the Mn $t_{2d}$ state is above the N
  $t_{2p}$ state in both spin channels and because the holes are in the Mn $d$
  states in GaMnN, the $p$-$d$ repulsion only affects the energy position of the
  $t_{2d}$ level, but has no direct effects on the stabilization of the FM phase [15].
  }
\end{thebibliography}
\end{document}